\documentclass[reprint,twocolumn,superscriptaddress,showpacs,preprintnumbers,nofootinbib,aps,prd,
]{revtex4-1}

\usepackage{graphicx,color}
\usepackage{amsmath,amssymb,amsfonts}
\usepackage{stackrel}
\usepackage{enumitem}
\usepackage[english]{babel}
\usepackage{verbatim}
\usepackage{hyperref}
\usepackage{comment}
\usepackage{ulem}
\usepackage{xcolor}
\usepackage{subcaption}
\usepackage{dcolumn}
\usepackage{physics}
\usepackage{dsfont}
\usepackage{booktabs} 
\usepackage{float}

\definecolor{darkred}{rgb}{.7,.1,.1}
\usepackage{hyperref}
\hypersetup{
    pdfnewwindow=true,  
    colorlinks=true,     
    linkcolor=darkred,     
    citecolor=blue,      
    filecolor=blue,      
    urlcolor=blue        
}
 
\graphicspath{
  {./}
  {figures/}
}

\usepackage{setspace}

\newcommand{\tf}{\texorpdfstring}

\def\nn{\nonumber}
\newcommand{\slashed}{\slash \hspace{-0.18cm}}

\newcommand{\cevns}{CE$\nu$NS}

\definecolor{orange}{rgb}{1,0.5,0}
\definecolor{amethyst}{rgb}{0.6, 0.4, 0.8}
\definecolor{antiquefuchsia}{rgb}{0.57, 0.36, 0.51}
\definecolor{byzantine}{rgb}{0.74, 0.2, 0.64}
\definecolor{blue-violet}{rgb}{0.54, 0.17, 0.89}
\definecolor{cadmiumred}{rgb}{0.89, 0.0, 0.13}
\definecolor{brightcerulean}{rgb}{0.11, 0.67, 0.84}

\begin{abstract}
\noindent

We investigate constraints on neutrino non-standard interactions (NSIs) in the effective field theory framework, using data from the first measurement of solar $^8$B neutrinos via coherent elastic neutrino-nucleus scattering (CE$\nu$NS) in the PandaX-4T and XENONnT experiments and data from the COHERENT experiment. The impacts of neutrino NSIs on the CE$\nu$NS cross section and the matter effect in the propagation of solar neutrinos are included, while we obtain that the expected number of CE$\nu$NS events is more sensitive to neutrino NSIs appearing in the cross section.
Due to relatively large statistical uncertainties, the sensitivities of the PandaX-4T and XENONnT experiments to the neutrino NSIs are currently limited, compared to the COHERENT experiment. Besides, we find that since the central value of the measured CE$\nu$NS counts significantly differs from the Standard Model prediction, the sensitivity of PandaX-4T experiment is even more restricted compared to XENONnT.
However, the measurements of PandaX-4T and XENONnT are uniquely sensitive to the neutrino NSIs for the $\tau$ flavor due to oscillation feature of the solar $^8$B neutrinos. 
We also assess how the experimental central value, exposure, and systematic uncertainties will affect the constraints on neutrino NSIs from various CE$\nu$NS measurements in the future.

\end{abstract}

\newcommand{\SunYATSET}{\affiliation{School of Physics and Astronomy, Sun Yat-sen University, Zhuhai 519082, P.R. China}}
\newcommand{\HIAS}{\affiliation{School of Fundamental Physics and Mathematical Sciences, Hangzhou Institute for Advanced Study, UCAS, Hangzhou 310024, China}}
\newcommand{\UCAS}{\affiliation{School of Physical Sciences, University of Chinese Academy of Sciences,   Beijing 100049, P.R. China}}
\newcommand{\ITP}{\affiliation{Institute of Theoretical Physics, Chinese Academy of Sciences,   Beijing 100190, P. R. China}}
\newcommand{\ICTPAP}{\affiliation{International Centre for Theoretical Physics Asia-Pacific, Beijing/Hangzhou, China}}

\begin{document}

\title{Constraints on neutrino non-standard interactions from  COHERENT, PandaX-4T and XENONnT}

\HIAS

\author{Gang Li}
\email{ligang65@mail.sysu.edu.cn}
\SunYATSET

\author{Chuan-Qiang Song}
\email{songchuanqiang21@mails.ucas.ac.cn}
\HIAS
\UCAS
\ITP

\author{Feng-Jie Tang}
\email{tangfengjie22@mails.ucas.ac.cn}
\HIAS
\UCAS
\ITP

\author{Jiang-Hao Yu}
\email{jhyu@itp.ac.cn}
\HIAS
\UCAS
\ITP
\ICTPAP

\maketitle

\section{Introduction}

In the Standard Model (SM), neutrinos interact with ordinary matter through the exchange of $W$ and $Z$ bosons.
Neutrino non-standard interactions (NSIs) in the charged and neutral currents beyond the SM (BSM) were initially formulated by Lee-Yang~\cite{Lee:1956qn} and Wolfenstein~\cite{Wolfenstein:1977ue}, respectively. 
While the detection of neutrinos in neutrino oscillation experiments only involves charged-current interactions, the coherent elastic neutrino-nucleus scattering (CE$\nu$NS)~\cite{Freedman:1973yd} serves as a unique probe of the neutral-current neutrino NSIs, potentially arising from new mediators, such as $Z^\prime$ boson~\cite{Barranco:2005yy}.

Despite the coherent enhancement of the \cevns\ cross section, this process is difficult to detect due to small deposited energy. It was first observed in the COHERENT experiment using the CsI[Na] scintillation detector~\cite{COHERENT:2017ipa} and later argon~\cite{COHERENT:2020iec} and germanium~\cite{Adamski:2024yqt} detectors with neutrinos produced from the spallation neutron source (SNS). These results as well as the follow-up detection with a larger exposure of CsI[Na]~\cite{COHERENT:2021xmm} and other experimental efforts~\cite{CONNIE:2019swq,CONUS:2020skt,nGeN:2022uje,Colaresi:2022obx,Su:2023klh,Ricochet:2023yek,Ackermann:2024kxo,Yang:2024exl,Cai:2024bpv,CICENNS}  have motivated diverse phenomenological studies~\cite{Coloma:2017ncl,Liao:2017uzy,Cadeddu:2017etk,Ge:2017mcq,AristizabalSierra:2017joc,Ciuffoli:2018qem,Farzan:2018gtr,Billard:2018jnl,AristizabalSierra:2018eqm,Cadeddu:2018izq,Altmannshofer:2018xyo,Miranda:2019skf,AristizabalSierra:2019ufd,Papoulias:2019txv,Giunti:2019xpr,Canas:2019fjw,Hoferichter:2020osn,Skiba:2020msb,Cadeddu:2020nbr,Du:2021rdg,Dasgupta:2021fpn,AtzoriCorona:2022qrf,DeRomeri:2022twg,Li:2024gbw}, see also Refs.~\cite{Lindner:2016wff,Dent:2016wcr,Coloma:2017egw} for earlier studies and Ref.~\cite{Abdullah:2022zue} for a recent review.

On the other hand, with the tremendous progress in the sensitivity of dark matter (DM) direct detection, it is anticipated that the DM experiments can detect neutrinos from astrophysical sources. These neutrinos exhibit nuclear recoil signatures resembling those of DM, which pose as irreducible backgrounds in DM direct detection and are often referred to as the ``neutrino fog"~\cite{Billard:2013qya,OHare:2021utq,Tang:2023xub,Tang:2024prl}. This is not unexpected since the idea of detecting DM that scatters off nuclei~\cite{Goodman:1985} was inspired by the proposal to detect MeV-range neutrinos via \cevns~\cite{Drukier:1984vhf}. The impact of neutrino NSIs on the limits of DM-nucleus scattering cross section in DM direct experiments has been studied in the literature, e.g.~\cite{AristizabalSierra:2017joc,Gonzalez-Garcia:2018dep,Chao:2019pyh}.

Recently, the solar ${^8}$B neutrino 
was measured through \cevns\ in the PandaX-4T~\cite{PandaX:2024muv} and XENONnT~\cite{XENON:2024ijk} experiments with corresponding statistical significance of $2.64\sigma$ and $2.73\sigma$, which signify the first step into the neutrino fog experimentally. Assuming the SM cross section of CE$\nu$NS, the signals are interpreted as the measurements of solar ${^8}$B neutrino flux of $(8.4\pm3.1)\times10^{6}\mathrm{~cm}^{-2} \mathrm{~s}^{-1}$ and $\left(4.7_{-2.3}^{+3.6}\right) \times 10^6 \mathrm{~cm}^{-2} \mathrm{~s}^{-1}$, respectively, both of which are consistent with the standard solar model predictions~\cite{Vinyoles:2016djt,Bahcall:1996qv,Bahcall:2005va} and the results from dedicated solar neutrino experiments~\cite{SNO:2011hxd,KamLAND:2011fld,Super-Kamiokande:2016yck,Borexino:2017uhp}.

In the SM, empirical nuclear form factors~\cite{Helm:1956zz, Klein:1999qj} that parameterize the nuclear response are usually adopted to calculate the cross section of CE$\nu$NS. However, an improved treatment is necessary if neutrino NSIs are present, which is feasible in the effective field theory (EFT) approach~\cite{Altmannshofer:2018xyo,Skiba:2020msb,Hoferichter:2020osn}\footnote{For vector and axial-vector NSIs, one can also refine the \cevns\ cross section by modifying the weak charge~\cite{Barranco:2005yy,Abdullah:2022zue}.
},
analogous to the situation of DM-nucleus scattering~\cite{Cirigliano:2012pq,Menendez:2012tm,Klos:2013rwa,Vietze:2014vsa,Hoferichter:2015ipa,Fitzpatrick:2012ix,Anand:2013yka,Bishara:2016hek,Bishara:2017pfq}.

An end-to-end EFT framework~\cite{Altmannshofer:2018xyo} was developed and utilized to describe the \cevns\ process from the new physics scale to the nuclear scale, which takes advantages of the heavy baryon chiral perturbation theory (HBChPT)~\cite{Jenkins:1990jv} and multipole expansions for nuclear responses~\cite{Walecka:1995mi}\footnote{Indeed, the multipole analysis was applied to neutrino-nucleus scattering processes in the SM weak charged currents slightly before the proposal of \cevns~\cite{OConnell:1972edu}. The effect of SM weak neutral currents in neutrino scattering off nuclei was later investigated in Ref.~\cite{Donnelly:1978tz}.
}. Compared to the nuclear form factor approach, the EFTs enable controlled uncertainties in systematic power countings, which allows for potential theoretical improvements in nuclear shell-model calculations~\cite{Hoferichter:2020osn} and beyond~\cite{AbdelKhaleq:2024hir}.

In this work, we will investigate the sensitivities to neutrino NSIs in the EFT framework, including matching between several EFTs. We consider relevant dimension-6 operators in the low-energy effective field (LEFT) and QCD chiral Lagrangian with external sources and heavy baryon expansion, and finally match to nuclear response function to obtain the most stringent constraints from the measurement conducted with the CsI[Na] detector in the COHERENT experiment~\cite{COHERENT:2021xmm} and the first results of new physics using the measurement of solar $^8$B neutrinos by PandaX-4T~\cite{PandaX:2024muv} and XENONnT~\cite{XENON:2024ijk}.

The remainder of the paper is organized as follows. In Sec.~\ref{sec:LEFT}, we discuss the neutrino NSIs from quark level to the nucleon level using the LEFT and HBChPT.
In Sec.~\ref{sec:COHERENTandWillson}, we derive the \cevns\ cross section using the multipole expansions for nuclear responses. In Sec.~\ref{sec:EventRates}, we evaluate the event rates of the \cevns\ signals in the COHERENT, PandaX-4T and XENONnT experiments,
and obtain the constraints on the Wilson coefficients for specific neutrino flavors, which are interpreted as lower bounds on the NSI energy scale using the $\chi^2$ analysis. Two-dimensional constraint on the NSI parameters is also obtained for the comparison of these two \cevns\ experiments. We have an assessment of the sensitivities of future measurements of solar $^8$B neutrinos via \cevns\ in DM detectors.
We conclude in Sec.~\ref{sec:conclusion}. In the appendices, we provide details of the detector resolution and efficiency in COHERENT CsI measurement, and the impact of NSI on the matter effect in the propagation of solar neutrinos.

\section{Neutrino non-standard interactions}
\label{sec:LEFT}

The neutral-current (NC) neutrino-quark interactions
can be parameterized as~\cite{Wolfenstein:1977ue,Scholberg:2005qs,Barranco:2005yy,Davidson:2003ha,Du:2021rdg}
\begin{align}
\label{eq:NSI-epsilon}
    \mathcal{L}_{\mathrm{NC}} &\supset - 2 \sqrt{2} G_F\left[ \epsilon_{\alpha \beta}^{q L}\left(\bar{\nu}_\alpha \gamma^\mu P_L \nu_\beta\right)\left(\bar{q} \gamma_\mu P_L q\right) \right. \nn \\
    &\quad \left.+\epsilon_{\alpha \beta}^{q R}\left(\bar{\nu}_\alpha \gamma^\mu P_L \nu_\beta\right)\left(\bar{q} \gamma_\mu P_R q\right)\right]\;,
\end{align}
where $P_{L/R} = (1\mp \gamma_5)/2$, $\alpha,\beta$ denote the flavors of neutrinos, and $q=u,d$, $G_F$ is the Fermi constant.

In the LEFT, the relevant effective Lagrangian is
\begin{equation}
\label{eq:LEFT}
    \mathcal{L}_{\rm LEFT} \supset \hat{\cal C}_{1,q}^{(6)} {\cal O}_{1,q}^{(6)} + \hat{\cal C}_{2,q}^{(6)} {\cal O}_{2,q}^{(6)}  \,,
\end{equation}
where the dimension-6 operators are defined as~\cite{Jenkins:2017jig,Altmannshofer:2018xyo}
\begin{align} 
{\cal O}_{1,q}^{(6)} & = (\bar \nu_\alpha \gamma_\mu P_L \nu_\beta) (\bar q \gamma^\mu q),\notag\\
{\cal O}_{2,q}^{(6)} & = (\bar \nu_\alpha \gamma_\mu P_L \nu_\beta)(\bar q \gamma^\mu \gamma_5 q)\,.
\end{align}

The correspondence between the coefficients $\epsilon_{\alpha\beta}^{qL(R)}$ and the Wilson coefficients $\hat{\cal C}_{1,q}^{(6)}$ and $\hat{\cal C}_{2,q}^{(6)}$ is 
\begin{align} \label{eq:epsLR}
    \epsilon_{\alpha\beta}^{qL/R}&= -\dfrac{1}{2\sqrt{2} G_F} \left( \hat{\mathcal C}_{1,q}^{(6)} \mp \hat{\mathcal C}_{2,q}^{(6)} \right)\;.
\end{align}

The SM contributions to the  Wilson coefficients after integrating out the $Z$ boson are~\cite{Altmannshofer:2018xyo,Abdullah:2022zue}
\begin{align}
    \left.\hat{\mathcal{C}}_{1, u(d)}^{(6)}\right|_{\mathrm{SM}}&=\mp \frac{G_F}{\sqrt{2}}\left(1-\frac{8(4)}{3} \sin^2\theta_W\right) \delta_{\alpha \beta}\;,\\
    \left.\hat{\mathcal{C}}_{2, u(d)}^{(6)}\right|_{\mathrm{SM}}&= \pm \frac{G_F}{\sqrt{2}} \delta_{\alpha \beta}\;.
\end{align}
where $\theta_W$ is the weak mixing angle with $\sin^2\theta_W = 0.2312$ \cite{ParticleDataGroup:2022pth}. We consider contributions from neutrino NSIs, so that 
\begin{align}
\label{eq:WC-SM_NSI}
    \hat{\mathcal{C}}_{i}^{(6)} = \left.\hat{\mathcal{C}}_{i}^{(6)}\right|_{\rm SM} + \left.\hat{\mathcal{C}}_{i}^{(6)}\right|_{\rm NSI}\;.
\end{align}
The dimensionful Wilson coefficient can also be expressed as 
\begin{align}
\label{eq:WC-NSI}
    \left.\hat{\mathcal{C}}_{i}^{(6)}\right|_{\rm NSI} = \dfrac{1}{\Lambda_{\rm NSI}^2} \left.\mathcal{C}_{i}^{(6)}\right|_{\rm NSI}\;,
\end{align}
where $\left.\mathcal{C}_{i}^{(6)}\right|_{\rm NSI}$ is dimensionless, and the energy scale $\Lambda_{\rm NSI}$ is determined by the mass of mediator that is responsible for the neutrino NSIs\footnote{For light mediator, there is additional momentum dependence from its propagator.}.

To obtain the neutrino-nucleus cross section of \cevns, we need to consider the matching in two steps~\cite{Altmannshofer:2018xyo,Hoferichter:2020osn}: (1) from the quark level to the nucleon level; (2) from the nucleon level to the nucleus level. In the first step, the nucleons in the target are considered non-relativistic due to the small momentum exchange $q$ compared to the nucleon mass. 
The interaction Lagrangian for the neutrinos and non-relativistic nucleons is
\begin{align}
    \mathcal{L}_{\rm NR} = c_{i,N}^{(d)} \mathcal{O}_{i,N}^{(d)} \;,
\end{align}
where $N=p,n$, and $d$ denotes the number of derivatives in the operator.

By using the HBChPT, the  following neutrino-nucleon operators at leading order are obtained:
\begin{align}
    \mathcal{O}_{1,N}^{(0)} & = (\bar \nu_\beta \gamma_\mu P_L \nu_\alpha) (v^\mu \bar N_v N_v)\,,\\
    \mathcal{O}_{2,N}^{(0)} & = (\bar \nu_\beta \gamma_\mu P_L \nu_\alpha)(\bar N_v S^\mu N_v)\,.
\end{align}
where $N_v$ denotes the large component of the nucleon field, and $v^\mu$ is the nucleon velocity, the spin operator $S^\mu = \gamma_5 \gamma_\perp^\mu /2$ with $\gamma_\perp^\mu \equiv \gamma^\mu - v^\mu \slashed v$.
In the lab frame, $v^\mu=(1,\vec 0)$, and $S^\mu = (0,\vec \sigma/2)$ with $\vec \sigma$ being the Pauli matrices. 
Denoting $p_1$ and $p_2$ ($k_1$ and $k_2$) as the momenta of income and outcome neutrinos (nucleons), respectively, we can define the momentum transfer as $q = p_{1} - p_{2}$.

From the quark-level interactions to the nucleon-level interactions, the matching conditions are expressed as\footnote{Note that we use the symbol $q$ to represent both the momentum transfer and the quark in the conventional manner.  }
\begin{align} \label{eq:c1(0)}
c_{1,N}^{(0)} & = \sum_{q=u,d} F_1^{q/N}  \hat {\cal C}_{1,q}^{(6)} \,,
&
c_{2,N}^{(0)} & =  2 \sum_{q=u,d} F_A^{q/N} \hat {\cal C}_{2,q}^{(6)}\,,
\end{align}
where $F_i $ denote the momentum-dependent nucleon form factors describing the hadronization of quark currents. We use the values of $F_i$ evaluated at $q^2 \to 0$~\cite{Bishara:2017pfq}, which are accurate enough for our purpose, 
\begin{align}
    F_1^{u/p} &= 2\;,& F_1^{d/p}&=1\;, \nn\\
    F_A^{u/p} &= 0.897\;,& F_A^{d/p}&=-0.376\;.
\end{align}

\section{Cross section of \tf{\cevns}{CEvNS}}
\label{sec:COHERENTandWillson}

In the second step, the nuclear response to neutrino scattering needs to be considered at the nuclear level, which is described similarly to DM detection~\cite{Fitzpatrick:2012ix,Anand:2013yka}. In this framework,
the many-body nuclear matrix elements are expanded using the multipole expansions~\cite{Walecka:1995mi}
in the harmonic oscillator basis, and can be calculated in the nuclear shell model \cite{Haxton:2008zza,Hoferichter:2020osn}.

To this end, we classify the Lagrangian terms according to charge operator ($1_N$) and nuclear spin operator ($\vec{S}$)~\cite{Altmannshofer:2018xyo}: 
\begin{equation} \label{eq:LNR}
\begin{aligned}
    \mathcal{L}_{\rm NR} &= \big( \bar{\nu}_{\alpha} l_{0,N} P_L \nu_{\beta} \big) 1_N  + \big( \bar{\nu}_{\alpha} \vec{l}_{5,N} P_L \nu_{\beta} \big) \cdot (2 \vec{S}) \,,
\end{aligned}
\end{equation}
where the Dirac structures are given by
\begin{align}
l_{0,N} =  c_{1,N}^{(0)} \slashed{v} \,, \quad
l_{5,N}^\mu = \frac{1}{2}c_{2,N}^{(0)} \gamma^\mu\;,
\end{align}
and $\vec l_5$ in Eq.~\eqref{eq:LNR} is the spatial three-vector components of $l_5^\mu$.

The differential cross section in the rest frame of the target nucleus is~\cite{Altmannshofer:2018xyo}\footnote{We have corrected a missing factor of $1/8$ in Eq.(3.23) of Ref.~\cite{Altmannshofer:2018xyo}, and have verified it by comparing the SM result with the calculation using the nuclear form factor~\cite{Lindner:2016wff,Abdullah:2022zue}.} 
\begin{equation}
\label{eq:xsec}
\frac{\rm{d} \sigma }{{\rm d} T_{\rm nr}}
=\frac{M}{ 8\pi  E_{\nu}^{2}}|\overline{{\cal M}}|_{RW}^{2}\;,
\end{equation}
where $M$ is the target nucleus mass, $E_\nu$ is the initial neutrino energy, $T_{\rm nr}$ is the nuclear recoil energy, and the spin-averaged amplitude square is expressed as
\begin{equation} \label{eq:MRW}
\begin{aligned}
|\overline{{\cal M}}|_{RW}^{2} = \frac{4\pi}{2J_A + 1} \sum_{\tau , \tau'=0,1} \Big(  R_{M}^{\tau\tau'} W_M^{\tau\tau'} \\
+ R_{\Sigma''}^{\tau\tau'}W_{\Sigma''}^{\tau\tau'}+ R_{\Sigma'}^{\tau\tau'}W_{\Sigma'}^{\tau\tau'}  \Big)\;.
\end{aligned}
\end{equation}
Here, $J_A$ is the spin of the target nucleus, 
$W_i$ denotes the nucleus response functions~\cite{Fitzpatrick:2012ix,Anand:2013yka}, and the kinematic factors $R_i$ are given by~\cite{Altmannshofer:2018xyo}
\begin{align} 
\begin{split}
R_{M}^{\tau \tau'}&= \mathrm{Tr} \big( P_L \slashed p_1  \gamma_0 l_{0,\tau'}^\dagger  \gamma_0\slashed p_2 l_{0,\tau}   \big) \,,
\end{split} 
\\
\begin{split}
R_{\Sigma''}^{\tau \tau'}&= \mathrm{Tr} \big( P_L \slashed p_1   \gamma_0 l_{5,\tau'}^{j\dagger}  \gamma_0 \slashed p_2 l_{5,\tau}^i \big) \hat{q}^i \,  \hat{q}^j \,,
\end{split}
 \\
R_{\Sigma'}^{\tau \tau'}&= \mathrm{Tr} \big( P_L \slashed{p}_1   \gamma_0 l_{5,\tau'}^{j\dagger}  \gamma_0 \slashed p_2 l_{5,\tau}^i  \big)\big(\delta^{ij}- \hat{q}^i \,  \hat{q}^j \big)\,,
\end{align}
where $\tau,\tau^\prime$ are the isospin indices, $\hat q \equiv \vec q/|\vec q|$, and $i,j=1,2,3$. In the isospin basis, $l_{0,\tau} = \left[l_{0,p} + (-1)^\tau l_{0,n} \right]/2$ and $l_{5,\tau}^\mu = \left[l_{5,p}^\mu + (-1)^\tau l_{5,n}^\mu \right]/2$.

\section{Event rates and constraints}
\label{sec:EventRates}
In the CE$\nu$NS experiments, neutrinos from the source will interact with detector target nuclei, causing nucleus recoils. The resulting signal can be translated into the event rate. 
In the following, we will investigate constraints on neutrino NSIs from the measurements by the COHERENT, PandaX-4T and XENONnT experiments.

\subsection{Constraints from COHERENT}
\label{ConstrainsCOHERENT}

We first consider the measurements of the  \cevns\ process in the COHERENT experiment using CsI[Na]~\cite{COHERENT:2021xmm}\footnote{We do not consider the COHERENT measurement using Ar detector~\cite{COHERENT:2020iec} since its sensitivity cannot compete with that CsI detector, even though the combination of these measurements can break degeneracy between different NSI parameter combinations~\cite{DeRomeri:2022twg}.
}.  
The differential event rate per target for neutrino flavor $\nu_\alpha=\nu_e,\nu_\mu,\bar\nu_\mu$ is expressed as~\cite{AbdelKhaleq:2024hir,Altmannshofer:2018xyo}:
\begin{equation}
    \frac{{{\rm d}} R_{\nu_\alpha}}{{{\rm d}} T_{\rm nr}} = \int_{E_{\nu, {\rm min}}}^{E_{\nu,{\rm max}}} dE_{\nu} \Phi_{\nu_\alpha}(E_{\nu}) \frac{{\rm d} \sigma}{{\rm d} T_{\rm nr}} \,,
\end{equation}
where ${{\rm d} \sigma}/{{\rm d} T_{\rm nr}}$ is the differential cross section
given in Eq.~\eqref{eq:xsec}.
The minimum initial neutrino energy is $E_{\nu,{\rm min}} \simeq \sqrt{MT_{\rm nr} /2} $, 
where $M$ is the nucleus mass. 
The upper integration limit is given by the maximal energy of initial neutrinos produced in $\pi^{+} \rightarrow \nu_{\mu} (\mu^{+} \rightarrow e^{+} \nu_{e} \bar{\nu}_{\mu})$.
For $\nu_e$ and $\bar \nu_\mu$, $E_{\nu,{\rm max}} = m_\mu/2 \simeq 52.8~ {\rm{MeV}}$, while for $\nu_\mu$, $E_{\nu,{\rm max}} = (m_\pi^2-m_\mu^2)/(2m_\pi) \simeq 30~ {\rm{MeV}}$~\cite{AtzoriCorona:2022qrf,AristizabalSierra:2018eqm}, where $m_\mu$ and $m_\pi$ are the mass of the muon and pion, respectively.

The total neutrino fluxes are described by the Michel spectrum~\cite{Coloma:2017egw,Liao:2017uzy}
\begin{align}
\Phi_{\nu_e}(E_\nu) &= {\cal N} \frac{192E_{\nu}^2}{m_\mu^3}\left(\frac{1}{2} - \frac{E_\nu}{m_\mu} \right)\,,
\\
\Phi_{\bar\nu_\mu}(E_\nu) &= {\cal N} \frac{64E_{\nu}^2}{m_\mu^3}\left(\frac{3}{4} - \frac{E_\nu}{m_\mu} \right)\,,
\\
\Phi_{\nu_\mu}(E_\nu) &= {\cal N}\delta\left(E_\nu - \frac{m_\pi^2 - m_\mu^2}{2 m_\pi} \right)\,,
\end{align}
where $\delta$ is the Dirac $\delta$-function, and the overall factor ${\cal N} \equiv r N_{\rm POT}/(4\pi L^{2})$ depends on the number of neutrinos $(r)$ that are produced for each proton on target (POT), the number of protons on target $(N_{\rm POT})$ and the distance between the source and the detector $(L)$. For the CsI[Na] detector in COHERENT experiment, $r=0.0848$, $N_{\rm POT} = 3.198 \times 10^{23}$ and $L = 19.3{~\rm m}$~\cite{COHERENT:2021xmm}. 

The time-integrated expected number of \cevns\ events in the $i$th bin of the number of photoelectrons (PEs) for the flavor $\nu_\alpha$ is given by~\cite{AristizabalSierra:2018eqm,Papoulias:2019txv,AtzoriCorona:2022qrf,DeRomeri:2022twg}
\begin{align}
\label{eq:event_coherent}
    N^{i}_{\nu_\alpha} &= n_N \sum_{x={\rm Cs, I}} \eta_x \langle \varepsilon_T \rangle_{\nu_\alpha}
    \int_{n_{\rm PE}^{i}}^{n_{\rm PE}^{i+1} } {\rm{d}} n_{\rm PE} \varepsilon\left(n_\mathrm{PE}\right) \nn\\
    &\quad \times \int_{T_{\rm nr, min}}^{T_{\rm nr, max}} {\rm d} T_{\rm nr}  P\left(n_{\rm PE} \right)   \left.\frac{{\rm{d}} R_{\nu_\alpha}}{{\rm{d}} T_{\rm nr}} \right|_x \,,
\end{align}
where we have taken into account the recoils of Cs and I with the fractions $\eta_{\rm Cs} = 51\%$ and $\eta_{\rm I} = 49\%$, respectively. 
The number of target nuclei in the detector is $n_{N} \equiv N_A M_{\rm det}/M_{T}$, where $M_{\rm det} = 14.6$~kg is the detector active mass, $M_{T} = 259.8~{\rm g/mol}$ is the molar mass of CsI, and $N_A$ denotes the Avogadro number. The detector energy resolution $P(n_{\rm PE})$ and efficiency $\varepsilon(n_{\rm PE})$ as well as the average time efficiency $\langle \varepsilon_T\rangle_{\nu_\alpha}$ are described in Appendix~\ref{app:coherent}.

In Fig.~\ref{fig:CsIEvents}, we compare the expected number of \cevns\ events in the SM as a function of $n_{\rm PE}$, which is calculated using the nuclear response functions described in Sec.~\ref{sec:COHERENTandWillson}, with the experimental data from COHERENT.  The contributions from different neutrino fluxes are included.

\begin{figure}
    \centering
    \includegraphics[width=0.8\linewidth]{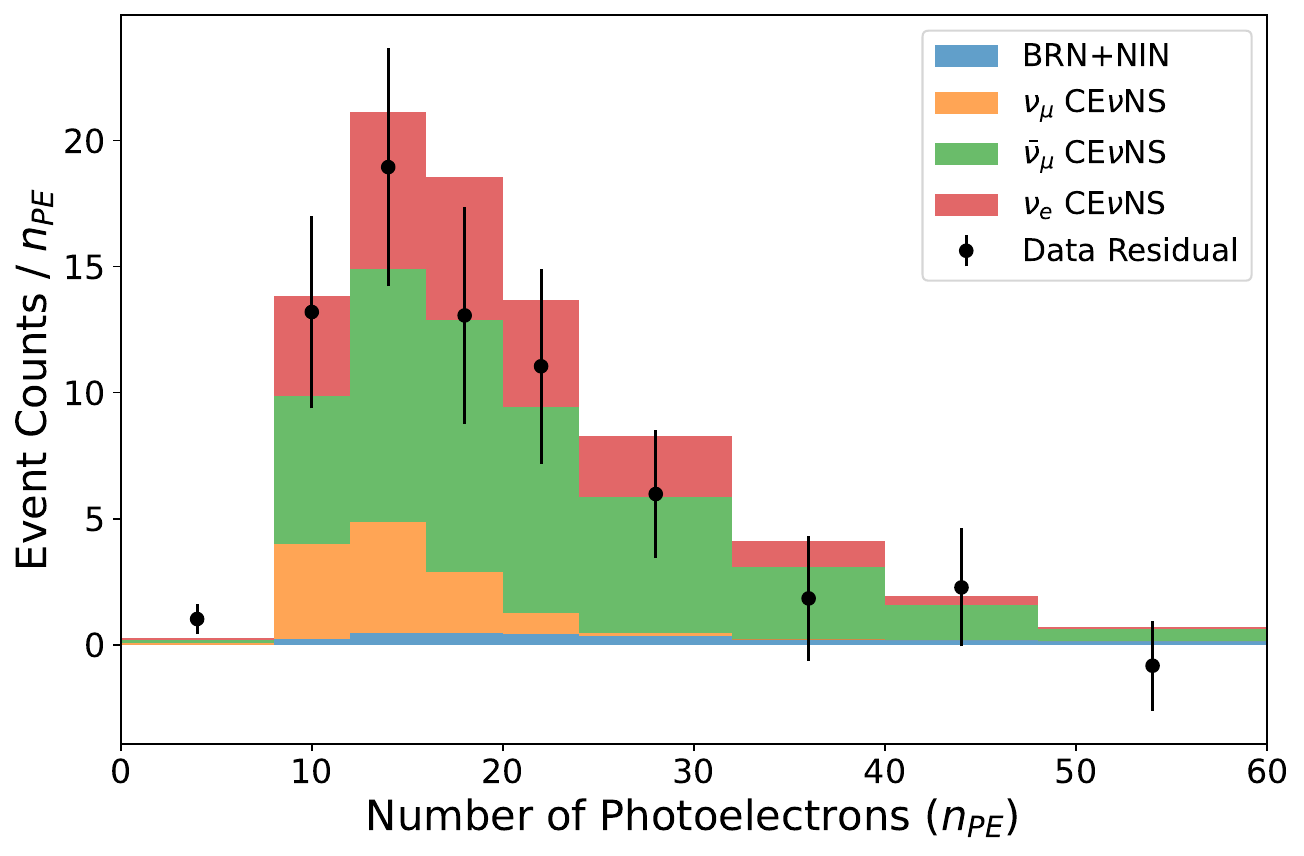}
    \caption{The comparison of the total expected event counts for \cevns\ events in the SM calculated using the nuclear response functions, with the experimental data (black points with error bars) collected with CsI[Na] detector by the COHERENT Collaboration~\cite{COHERENT:2021xmm}. The numbers of events from different flavors are shown.}
    \label{fig:CsIEvents}
\end{figure}

To constrain the neutrino NSIs, we perform the binned $\chi^2$ analysis using the following least-squares function~\cite{Fogli:2002pt,AristizabalSierra:2018eqm},
\begin{align}
\label{eq:chi2CsI}
    \chi^2 = &\sum_{i=1}^9 \frac{\left[N_{\rm meas}^{i} - N_{\text{CE}\nu\text{NS}}^{i} (1 + \alpha) - N_{\rm BRN+NIN}^{i} (1 + \beta) \right]^2}{(\sigma_{\rm stat}^{i})^2} \nn \\
    & + \left(\frac{\alpha}{\sigma_{\alpha}} \right)^2 + \left(\frac{\beta}{\sigma_{\beta}} \right)^2 \,,
\end{align}
where $N_{\rm meas}^i$ and  $N_{\rm BRN+NIN}^{i}$ represent the measured number of \cevns\ events, and the expected number of beam-related neutron (BRN) and neutrino-induced neutron (NIN) background events in the $i$th bin of $n_{\rm PE}$, respectively. The associated statistical uncertainty is $\sigma_{\rm stat}^{i} = \sqrt{N_{\rm meas}^{i} + N_{\rm BRN+NIN}^{i}} $. The expected number of \cevns\ events is given by 
\begin{align}
    N_{\text{CE}\nu\text{NS}}^i = N_{\nu_e}^i + N_{\nu_\mu}^i + N_{\bar \nu_\mu}^i\;,
\end{align}
which depends on the neutrino NSIs. 

The relative systematic uncertainties from the quenching factor $(3.8\%)$, neutrino flux $(10\%)$ and signal acceptance $(4.1\%)$~\cite{COHERENT:2021xmm,DeRomeri:2022twg}, and the response functions $(5\%)$ are considered\footnote{It is noted that in the approach of nuclear form factor, the relative systematic uncertainty is about $5\%$~\cite{Papoulias:2019txv}. Here, we assume that the uncertainties associated with the response functions are comparable.  
}, which lead to the total uncertainty $\sigma_\alpha=0.125$. Besides, $\sigma_\beta = \sqrt{\sigma_{\rm BRN}^2 + \sigma_{\rm NIN}^2}$ with the uncertainties of BRN and NIN backgrounds are $\sigma_{\rm BRN}=0.25$ and $\sigma_{\rm NIN}=0.35$, respectively.

The quantity $\chi^2$ is minimized over the systematic nuisance parameters $\alpha$ and $\beta$,
so that we can derive the 90\% confidence level (C.L.) bounds on the Wilson coefficients of neutrino NSIs by requiring $\Delta \chi^2 \equiv \chi^2 - \chi_{\rm min}^2 \leq 2.71$.
In Eq.~\eqref{eq:WC-NSI}, assuming $\left.\mathcal{C}_{i}^{(6)}\right|_{\rm NSI} =1$ and summing over the fluxes of $\nu_{\mu}$ and $\bar \nu_{\mu}$, we obtain the one-parameter-a-time lower bounds on $\Lambda_{\rm NSI}$ for specific flavors, which are presented in Table~\ref{tab:lambda bounds CsI}.

\begin{table}[!htb]
    \centering
    \begin{tabular}{ccc}
    \hline
    $\Lambda_{\rm NSI}/{\rm GeV}$ & $\nu_{e}$ & $\nu_{\mu}$  \\
    \hline
    $\hat{\mathcal C}_{1,u}^{(6)}$ & 390  & 395    \\
    $\hat{\mathcal C}_{1,d}^{(6)}$ & 407  & 414    \\
    $\hat{\mathcal C}_{2,u}^{(6)}$ & 44.7   & 68.4   \\
    $\hat{\mathcal C}_{2,d}^{(6)}$ & 26.4   & 40.9   \\
    \hline
    \end{tabular}
    \caption{Lower bounds on $\Lambda_{\rm NSI}$ in units of GeV for the Wilson coefficients $\left.\hat{\mathcal{C}}_{i}^{(6)}\right|_{\rm NSI} = \hat{\mathcal{C}}_{i}^{(6)} -\left.\hat{\mathcal{C}}_{i}^{(6)}\right|_{\rm SM}$ from COHERENT~\cite{COHERENT:2021xmm}.} 
    \label{tab:lambda bounds CsI}
\end{table}

\subsection{Constraints from measurements of solar ${^8}$B neutrinos}
\label{sec:Constrains8B}

The measurements of solar ${^8}$B neutrinos in the \cevns\ process~\cite{,PandaX:2022aac,PandaX:2024muv,XENON:2024ijk} can also impose constraints on the neutrino NSIs. 
In this work, we consider the recent results of the PandaX-4T~\cite{PandaX:2024muv} and XENONnT~\cite{XENON:2024ijk} experiments using the liquid xenon.

The differential event rate per target for neutrino flavor $\nu_\alpha=\nu_e,\nu_\mu,\nu_\tau$ is expressed as
\begin{equation} \label{eq.diffEventRate8B}
    \frac{{{\rm d}} R_{\nu_\alpha}}{{{\rm d}} T_{\rm nr}} =  \int_{E_{\nu, {\rm min}}}^{E_{\nu,{\rm max}}} dE_{\nu}  \Phi_{\nu_\alpha}(E_{\nu})  \frac{{\rm d} \sigma}{{\rm d} T_{\rm nr}} \,,
\end{equation}
where the minimum neutrino energy $E_{\nu,\rm min}\simeq \sqrt{M T_{\rm nr} /2}$ with $M$ the mass of $^{131}$Xe, and the maximum energy $E_{\nu,\rm max}$ is about 16 $\rm MeV$ \cite{Bahcall:1996qv}.

The solar $^8$B neutrino $\nu_e$ is produced in the Sun, and then propagates to the Earth. 
The total flux of neutrino $\nu_\alpha$ detected at the Earth is defined as
\begin{align}
\label{eq:flux_B8}
    \Phi_{\nu_\alpha}(E_{\nu})  = \dfrac{\mathcal{E}}{M_{\rm det}} \langle P_{\nu_\alpha} \rangle  \phi(^8{\rm B})\;,
\end{align}
where $\mathcal{E} $ and $M_{\rm det}$ 
denote the exposure and detector active mass, respectively, and $\phi(^8{\rm B}) = 5.46(1 \pm 0.12) \times 10^6 ~{\rm cm^{-2}s^{-1}}$ is the predicted solar $^8$B neutrino flux~\cite{Vinyoles:2016djt}\footnote{The predictions by the other groups based on the standard solar model can be found in Refs.~\cite{Bahcall:1996qv,Bahcall:2005va}. 
}. $\langle P_{\nu_\alpha} \rangle$ is the probability of solar neutrino $\nu_e$ to manifest as $\nu_\alpha$ at the Earth averaged over the exposure.

\begin{figure}[!htb]
    \centering
    \includegraphics[width=0.8\linewidth]{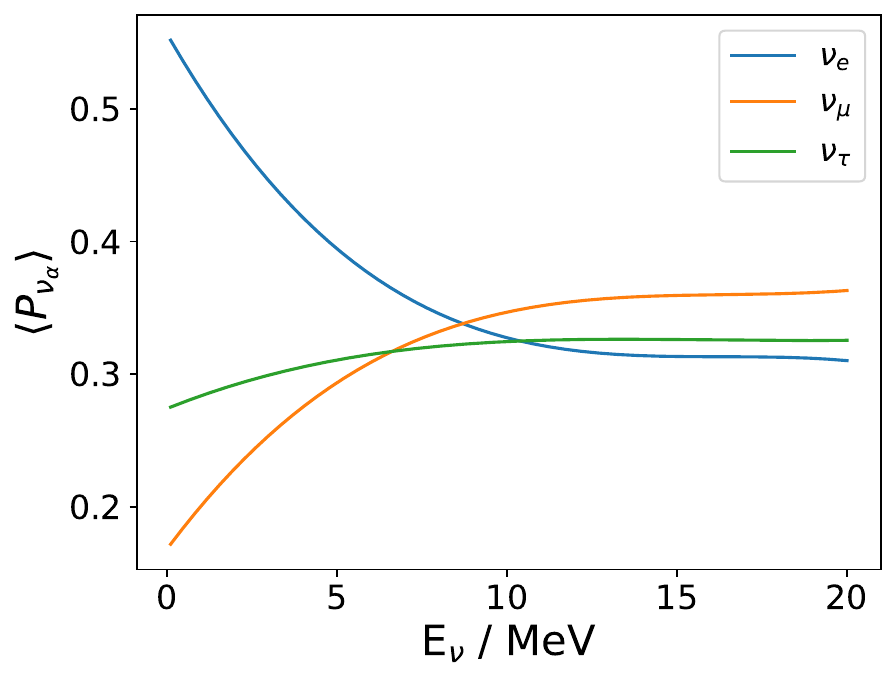}
    \caption{The probabilities of various neutrino fluxes at the Earth averaged over the exposure period, which are computed with the package PEANUTS~\cite{Gonzalo:2023mdh}.}
    \label{fig:8BProbability}
\end{figure}

Due to the neutrino oscillation, the flavor composition of solar neutrinos detected at the Earth differs from that produced in the Sun.
Figure~\ref{fig:8BProbability} shows the averaged probability $\langle P_{\nu_\alpha} \rangle$ for different flavors of neutrinos with matter effects in the neutrino propagation being included, which is computed with the package PEANUTS~\cite{Gonzalo:2023mdh}.
 
In the full analysis, we have included the impact of neutrino NSIs on the solar matter effect in the propagation, the details of which are given in Appendix~\ref{app:propagation}.
 In Fig.~\ref{fig.averPee}, we illustrate the averaged survival probabilities of electron neutrino  $\langle P_{\nu_e} \rangle$ for different choices of the NSI parameter $\epsilon_{ee}^{uV}$, which are validated against PEANUTS. It is shown that the neutrino NSIs have substantial impact on the solar matter effect, leading to the averaged probabilities varying by a factor of 2 at most for $\left|\epsilon_{ee}^{uV}\right|\leq 0.5$.

\begin{figure}[!h]
    \centering
    \includegraphics[width=0.8\linewidth]{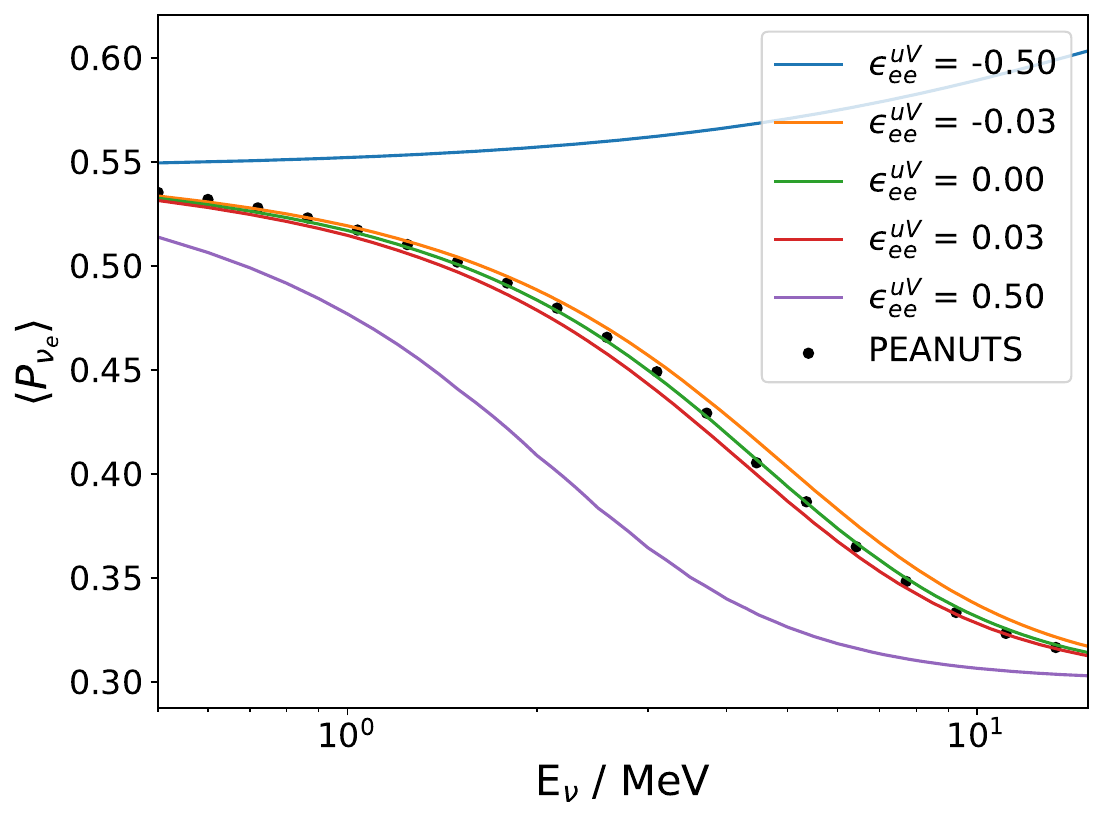}
    \caption{
    The averaged survival probabilities of electron neutrino for different choices of the NSI parameter $\epsilon_{ee}^{uV}$.
    }
    \label{fig.averPee}
\end{figure}

The expected number of \cevns\ events for the flavor $\nu_\alpha$ is given by~\cite{Li:2024gbw},
\begin{align}
    \label{eq:event_B8}
    N_{\nu_\alpha} = n_N  \int^{T_{\rm nr,max}}_{T_{\rm nr,min}} {\rm{d}} T_{\rm nr} ~ \varepsilon(T_{\rm nr})  \frac{{\rm{d}} R_\alpha}{{\rm{d}} T_{\rm nr}} \,,
\end{align}
where $\varepsilon(T_{\rm nr})$ is the detection efficiency, which depends on the nuclear recoil function. The number of target nuclei $(^{131}\rm Xe)$ in the detectors of the PandaX-4T and XENONnT experiments is $n_{N} \equiv N_A M_{\rm det}/M_{T}$ with the molar mass $M_{T} = 131.29~{\rm g/mol}$. 

To constrain the neutrino NSIs, we perform the single-bin $\chi^2$ analysis~\cite{Fogli:2002pt},
\begin{equation}
\label{eq:chi2}
    \chi^2 =\frac{\left[N_{\rm meas} - N_{\text{CE}\nu\text{NS}}(1 + \alpha) \right]^2}{\sigma_{\rm stat}^2} + \left(\frac{\alpha}{\sigma_{\alpha}} \right)^2  \,.
\end{equation}
Here, the expected number of \cevns\ events is given by 
\begin{align}
    N_{\text{CE}\nu\text{NS}} = N_{\nu_e} + N_{\nu_\mu} + N_{\nu_\tau}\;.
\end{align}

From the expressions in Eqs.~\eqref{eq.diffEventRate8B}~\eqref{eq:event_B8}, $N_{\text{CE}\nu\text{NS}}$ depends on the averaged probability and the \cevns\ cross section, both of which are affected by the neutrino NSIs. We will see in Fig.~\ref{fig:chiSquare8BCsI} that $N_{\text{CE}\nu\text{NS}}$ is more sensitive to neutrino NSIs appearing in the cross section, which is approximated as a quadratic polynomial of the NSI parameters. 

\subsubsection{PandaX-4T result}

Two datasets are collected by PandaX-4T, which differ in the energy threshold, as displayed in Table~\ref{tab:PandaX-4TData}. Note that the numbers of signal events for the pair data $(3.5)$ and US2 data $(75)$ are obtained from the combined likelihood fit~\cite{PandaX:2024muv}. Given the smaller number of signal events in paired data, we only consider the US2 data.
The exposure in Eq.~\eqref{eq:flux_B8} is thus given by $\mathcal{E} = 1.04~ {\rm tonne} \cdot {\rm year} $.

\begin{table}[h]
    \centering
    \begin{tabular}{ccc} 
    \hline
    PandaX-4T data & paired & US2  \\
    \hline
    energy threshold & $1.1$~keV & $0.33$~keV
    \\
    total exposure & $1.25$ tonne$\cdot$year & $1.04$ tonne$\cdot$year
    \\
    event number & 3.5 & 75
    \\
    \hline
    \end{tabular}
    \caption{The experimental data of detecting solar $^8$B neutrinos in the \cevns\ process by PandaX-4T~\cite{PandaX:2024muv}. The energy thresholds, total exposures, and numbers of signal events for the paired and US2 data are given.
    } 
    \label{tab:PandaX-4TData}
\end{table}

In the analysis by the PandaX-4T Collaboration~\cite{PandaX:2024muv}, the signals are interpreted in terms of the measured solar $^8$B neutrino flux assuming the SM cross section of \cevns, which is $(8.4\pm3.1)\times10^{6}\mathrm{~cm}^{-2}$ with the relative statistical uncertainty being $37\%$. Therefore, for our purpose, instead of taking the number of signal events post the combined likelihood fit, we calculate the number of signal events for the US2 data using the signal efficiency from Figure 1 of Ref.~\cite{PandaX:2024muv}. We obtain the measured number of \cevns\ events $N_{\rm meas} = 69.1$ with the statistic uncertainty $ \sigma_{\rm stat} = 0.37 \times N_{\rm meas}$.
The SM prediction is $\left.N_{\text{CE}\nu\text{NS}}\right|_{\rm SM} = 44.9$~\cite{Vinyoles:2016djt}, which implies that $N_{\rm meas} \simeq 1.54 \left.N_{\text{CE}\nu\text{NS}}\right|_{\rm SM}$ for the PandaX-4T measurement  using the US2 data.

The relative systematic uncertainties from the selection efficiency $(12\%)$, signal modeling $(17\%)$ and solar $^8$B neutrino flux $(12\%)$~\cite{PandaX:2024muv,Vinyoles:2016djt}, and the response functions $(5\%)$ are considered, which lead to the total uncertainty $\sigma_{\alpha} = 0.245 $. 

Again, $\chi^2$ is minimized over the nuisance parameter $\alpha$ to derive the 90\% C.L. limits on the Wilson coefficients by requiring $\Delta \chi^2 \equiv \chi^2 - \chi_{\rm min}^2 \leq 2.71$.  In Eq.~\eqref{eq:WC-NSI}, assuming $\left.\mathcal{C}_{i}^{(6)}\right|_{\rm NSI} =1$, we obtain the one-parameter-a-time lower bounds on $\Lambda_{\rm NSI}$ for specific flavors, which are presented in Table~\ref{tab:lambda bounds 8B}.

\begin{table}[h]
    \centering
    \begin{tabular}{cccc} 
    \hline
    $\Lambda_{\rm NSI}/{\rm GeV}$ & $\nu_{e}$ & $\nu_{\mu}$ & $\nu_{\tau}$ \\
    \hline
    $\hat{\mathcal C}_{1,u}^{(6)}$ & 287.46  & 289.61    & 286.70  \\
    $\hat{\mathcal C}_{1,d}^{(6)}$ & 304.61  & 306.88  	 & 303.80  \\
    $\hat{\mathcal C}_{2,u}^{(6)}$ & 14.70   & 14.81     & 14.60     \\
    $\hat{\mathcal C}_{2,d}^{(6)}$ & 23.32   & 23.48     & 23.15     \\
    \hline
    \end{tabular}
    \caption{Lower bounds on $\Lambda_{\rm NSI}$ in units of GeV for the Wilson coefficients $\left.\hat{\mathcal{C}}_{i}^{(6)}\right|_{\rm NSI} = \hat{\mathcal{C}}_{i}^{(6)} -\left.\hat{\mathcal{C}}_{i}^{(6)}\right|_{\rm SM}$ from PandaX-4T~\cite{PandaX:2024muv}. } 
    \label{tab:lambda bounds 8B}
\end{table}

Owing to the neutrino oscillation, a significant portion of the solar neutrino fluxes reaching the Earth are composed of $\nu_\tau$ as depicted in Fig.~\ref{fig:8BProbability}. Therefore, the measurements of solar ${^8}$B neutrinos via \cevns\ can give unique constraints on the neutrino NSIs for the $\tau$ flavor. For the $e$ and $\mu$ flavors, the measurement of COHERENT CsI provides more stringent constraints on the Wilson coefficients of NSIs.

\begin{figure}[!htb]
    \centering
    \includegraphics[width=0.8\linewidth]{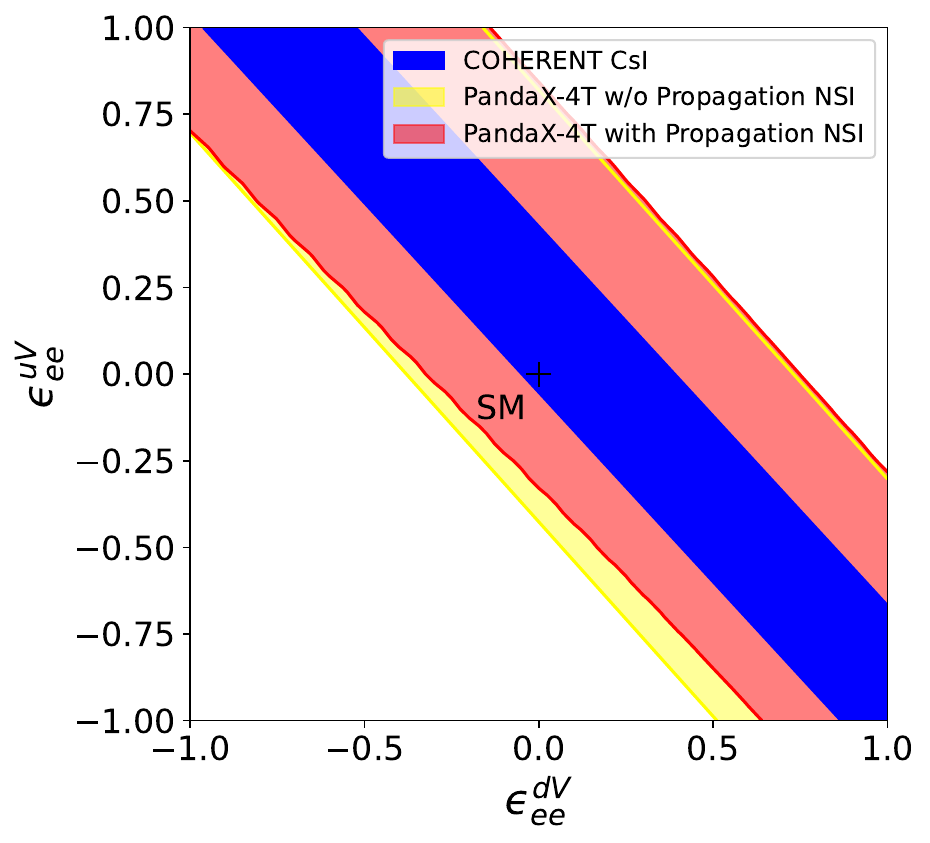}
    \caption{The 90\% C.L. allowed regions of $\epsilon_{ee}^{uV}$ and $\epsilon_{ee}^{dV}$ by the measurements in the COHERENT CsI~\cite{COHERENT:2021xmm} experiment and PandaX-4T~\cite{PandaX:2024muv} experiment  with or without considering the impact of the neutrino NSIs on solar matter effect in the propagation, as depicted in blue, red, and yellow colors, respectively. }
    \label{fig:chiSquare8BCsI}
\end{figure}

For comparison, we also obtain the two-dimensional constraint on the neutrino NSI parameters $\epsilon_{ee}^{uV}$ and $\epsilon_{ee}^{dV}$, which are defined as~\cite{Abdullah:2022zue}
\begin{align} \label{eq:epsee}
    \epsilon_{ee}^{qV} = \dfrac{-1}{\sqrt{2}G_F} \left. \hat{\mathcal{C}}_{1,q}^{(6)}\right|_{\rm NSI}\;.
\end{align}
By requiring $\Delta \chi^2 \leq 4.61$, we obtain the 90\% C.L. allowed regions in Fig.~\ref{fig:chiSquare8BCsI}.
Note that our fitted result using the COHERENT CsI measurement agrees with Ref.~\cite{DeRomeri:2022twg}. It is shown that the constraints on $\epsilon_{ee}^{uV}$ and $\epsilon_{ee}^{dV}$ from the \cevns\ measurement by PandaX-4T are weaker than those from COHERENT.

For the two-dimensional NSI analysis of the PandaX-4T measurement, we compare the results with  or without considering the impact of the neutrino NSIs on solar matter effect in the propagation, which leads to the change in the slope of the allowed region. We can see that such an impact is milder compared to the variation of the NSI parameters. This is because the expected number of \cevns\ events is more sensitive to neutrino NSIs appearing in the cross section, which is approximated as a quadratic polynomial of the NSI parameters.

\subsubsection{XENONnT result}

The XENONnT experiment has also measured the nuclear recoils from solar $^8$B neutrinos via \cevns\ process
with the liquid xenon target \cite{XENON:2024ijk}. The energy threshold in the XENONnT experiment is $0.5$~keV, and the total exposure is 3.51 tonne$\cdot$year. The event number of the $^8$B signal is $10.7\pm3.95$, where
$N_{\rm meas} = 10.7$ and $\sigma_{\rm stat} = 3.95$. The systematic uncertainties stem from the $^8$B solar neutrino flux ($12\%$) \cite{SNO:2011hxd}  and the response functions ($5\%$). and the quadratic sum gives the total systematic uncertainty $\sigma_\alpha = 13\%$. For the given exposure, we obtain the event number predicted in the SM  $\left.N_{\text{CE}\nu\text{NS}}\right|_{\rm SM} = 11.15$. Thus the central value  measured by XENONnT is only slightly smaller than the SM prediction.

\begin{figure}[!h]
    \centering    \includegraphics[width=0.8\linewidth]{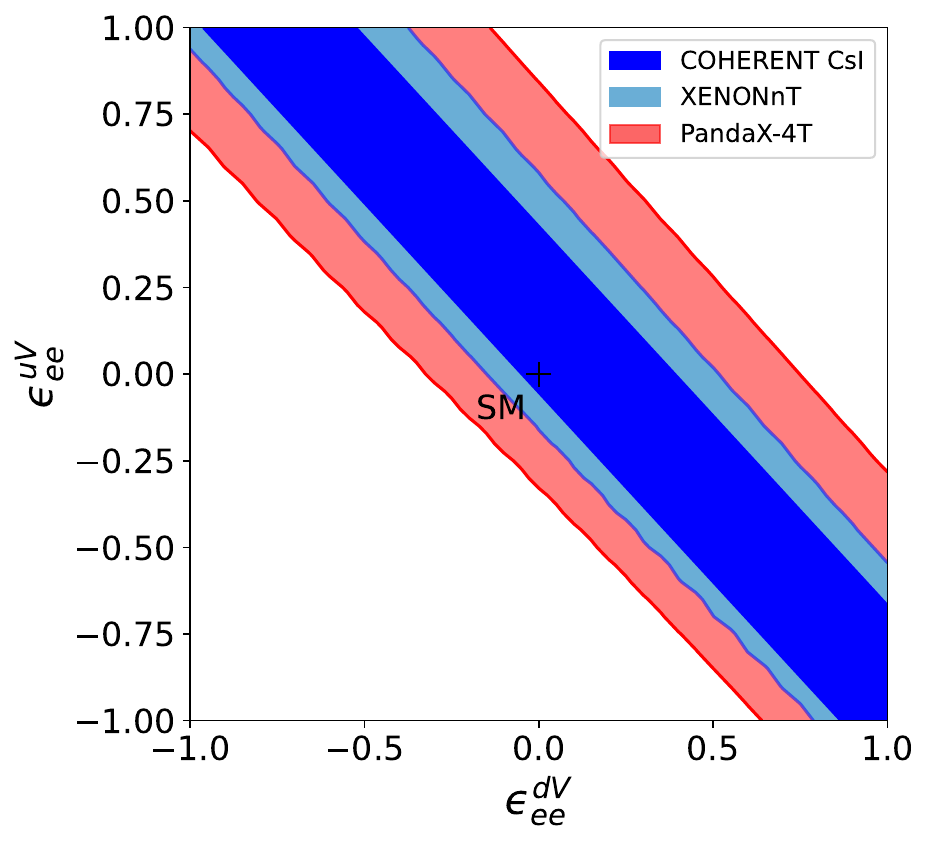}
    \caption{The 90\% C.L. allowed regions of $\epsilon_{ee}^{uV}$ and $\epsilon_{ee}^{dV}$ by the measurements in the COHERENT CsI (blue), PandaX-4T (red) and XENONnT (light blue) experiments 
    with the impact of neutrino NSIs on the solar matter effects in the propagation being included.
    }
    \label{fig:PandaXENONnT-NSI}
\end{figure}

We also perform the $\chi^2$ analysis of the XENONnT measurement.
The one-parameter-a-time lower bounds on the scale $\Lambda_{\rm NSI}$ using the data from XENONnT are presented in Table \ref{tab:lambda bounds 8B-XENONnT}.
In Fig.~\ref{fig:PandaXENONnT-NSI}, we compare the two-dimensional allowed regions of the NSI parameters $\epsilon_{ee}^{uV}$ and $\epsilon_{ee}^{dV}$ by the measurements in the COHERENT, PandaX-4T and XENONnT experiments. It is shown that the constraint from XENONnT is tighter than that from PandaX-4T, but it is still weaker than that from COHERENT due to larger uncertainties.

\begin{table}[h]
    \centering
    \begin{tabular}{cccc} 
    \hline
    $\Lambda_{\rm NSI}/{\rm GeV}$ & $\nu_{e}$ & $\nu_{\mu}$ & $\nu_{\tau}$ \\
    \hline
    $\hat{\mathcal C}_{1,u}^{(6)}$ &342.20 &344.97 &342.50 \\
    $\hat{\mathcal C}_{1,d}^{(6)}$ &388.46 &393.08 &388.95 \\
    $\hat{\mathcal C}_{2,u}^{(6)}$ &19.67 	&20.05 	&19.69 \\
    $\hat{\mathcal C}_{2,d}^{(6)}$ &31.19 	&31.80 	&31.23 \\
    \hline
    \end{tabular}
    \caption{Lower bounds on $\Lambda_{\rm NSI}$ in units of GeV for the Wilson coefficients $\left.\hat{\mathcal{C}}_{i}^{(6)}\right|_{\rm NSI} = \hat{\mathcal{C}}_{i}^{(6)} -\left.\hat{\mathcal{C}}_{i}^{(6)}\right|_{\rm SM}$ from XENONnT~\cite{XENON:2024ijk}. } 
    \label{tab:lambda bounds 8B-XENONnT}
\end{table}

\begin{figure}[!h]
    \centering
    \includegraphics[width=0.8\linewidth]{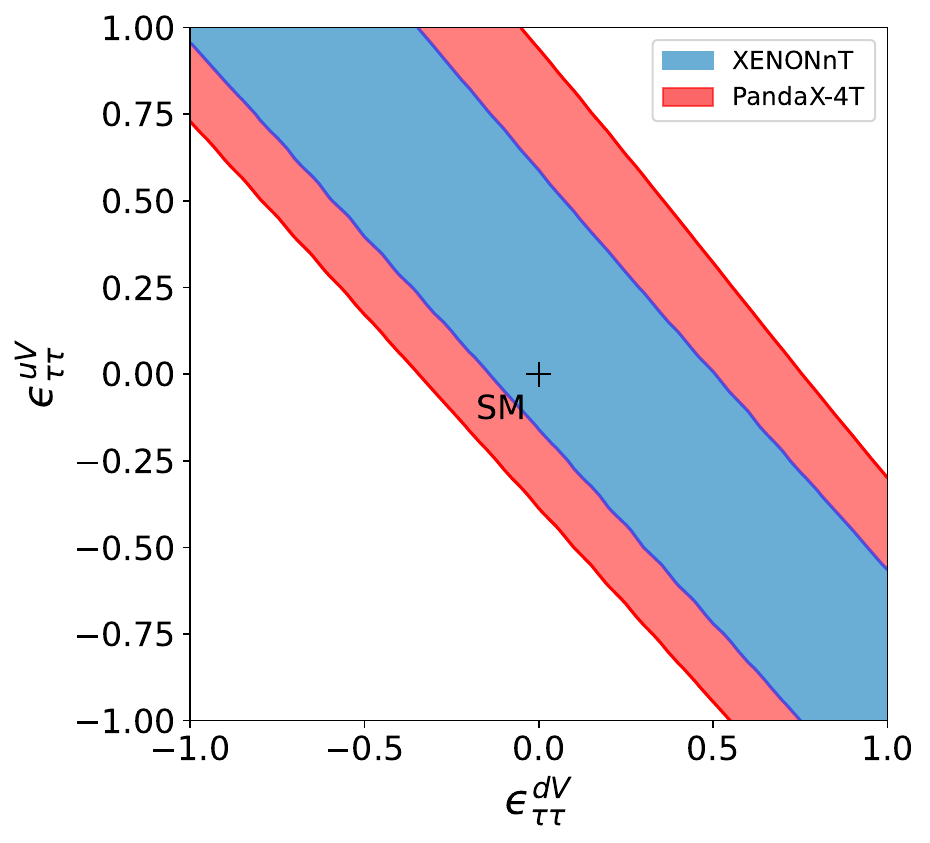}
    \caption{
    The 90\% C.L. allowed regions of $\epsilon_{\tau\tau}^{uV}$ and $\epsilon_{\tau\tau}^{dV}$ by the measurements in the PandaX-4T (red) and XENONnT (light blue) experiments with the impact of neutrino NSIs on the solar matter effects in the propagation being included.
    }
    \label{fig:chiSquareTau}
\end{figure}

As we have mentioned, the measurements of solar ${^8}$B neutrinos via the \cevns\ process in the PandaX-4T and XENONnT experiments can give unique constraints on the neutrino NSIs for the $\tau$ flavor. In Fig.~\ref{fig:chiSquareTau}, we show the two-dimensional allowed regions of the NSI parameters $\epsilon_{\tau\tau}^{uV}$ and $\epsilon_{\tau\tau}^{dV}$, which are defined similar to $\epsilon_{ee}^{qV}$ $(q=u,d)$ in Eq.~\eqref{eq:epsee} with $e$ being replaced by $\tau$. 

\section{Prospects for PandaX-4T}
\label{sec:Prospects}

\begin{figure}[h] 
    \centering
    \includegraphics[width=0.7\linewidth]{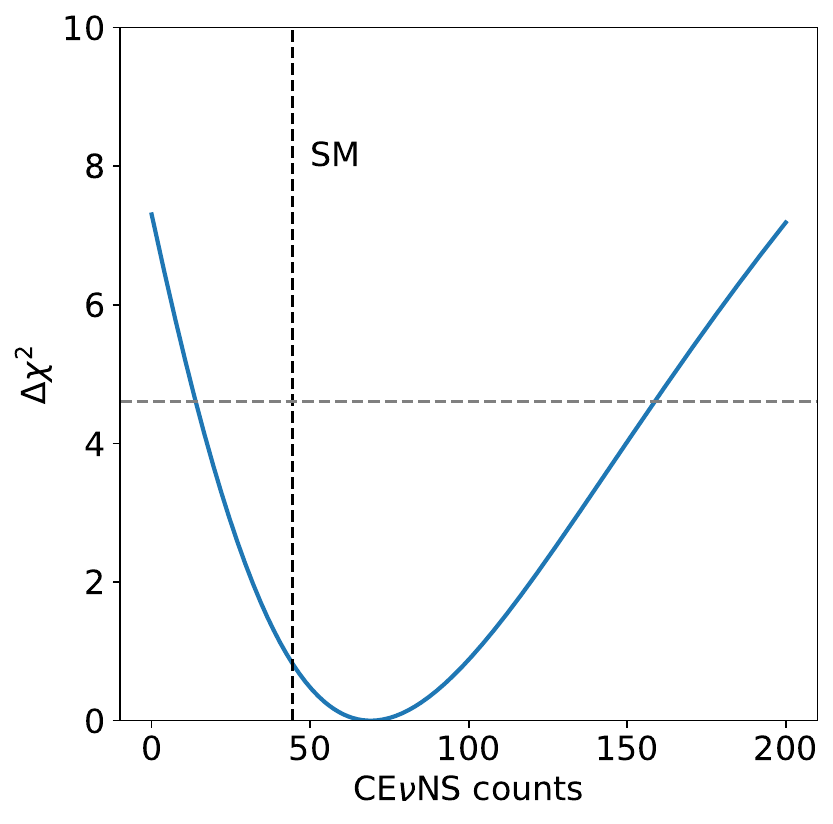}
    \caption{The distribution of $\Delta\chi^2$ as a function of the \cevns\ counts. The SM prediction of the \cevns\ counts is 44.9, located at the endpoint of the black dotted line. The central value of measured \cevns\ counts in the PandaX-4T experiment (US2 data) is $69.1$, located at the minimum of $\Delta\chi^2$ distribution. The gray line corresponds to $\Delta\chi^2 = 4.61$.
    }
    \label{fig:OneDimchiSquare}
\end{figure}

\begin{figure}[h]
    \centering
    \includegraphics[width=0.8\linewidth]{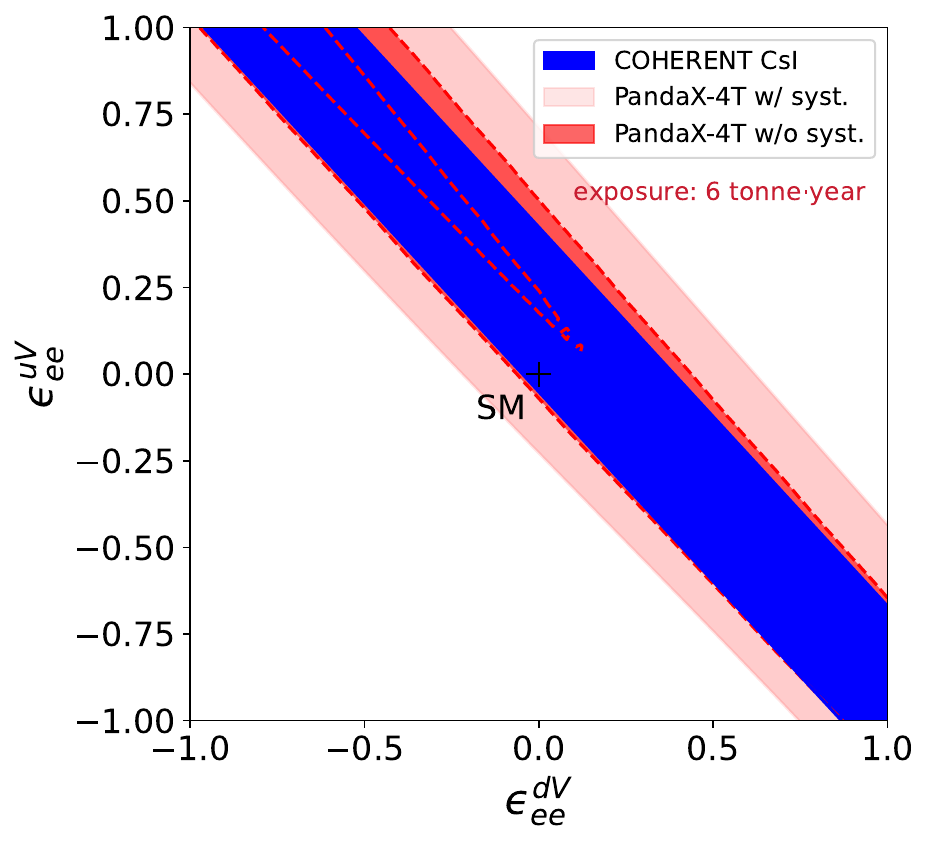}
    \caption{The 90\% C.L. allowed regions of $\epsilon_{ee}^{uV}$ and $\epsilon_{ee}^{dV}$ by the COHERENT CsI measurement (blue) and anticipated measurements by PandaX-4T assuming the experimental central value $N_{\rm meas} = \left.N_{\text{CE}\nu\text{NS}}\right|_{\rm SM} = 44.9$ and exposure $\mathcal{E} = 6~ {\rm tonne} \cdot {\rm year} $. 
     The lighter and darker red regions with solid and dashed boundaries are obtained with the systematic uncertainty $\sigma_\alpha = 0.245$ and 0, respectively. The smaller region with dashed red boundary is excluded. 
     }
    \label{fig:CentralValuechiSquare8BCsI}
\end{figure}

Apart from the larger uncertainties than the COHERENT measurement, we find that the deviation of the central value of experimentally measured \cevns\ counts has a significant impact on the sensitivities of PandaX-4T to the NSI parameters.
To understand it and make projections of the future prospects, we present the $\Delta\chi^2$ distribution as a function of the \cevns\ counts in Fig.~\ref{fig:OneDimchiSquare}, using the US2 data from PandaX-4T. The upper limit of the \cevns\ counts with $\Delta \chi^2\leq 4.61$ determines the boundaries of the blue bands in Fig.~\ref{fig:CentralValuechiSquare8BCsI}.  
It is verified that with the central value unchanged, the sensitivity of PandaX-4T shows a mild improvement for $\sim 5$ times larger exposure.

We further assess the sensitivities of future solar $^8$B neutrino measurements via \cevns\ in DM detectors, leveraging the current results from PandaX-4T as a reference.
In Fig.~\ref{fig:CentralValuechiSquare8BCsI}, we show the projected constraints on the NSI parameters under the assumption of $N_{\rm meas} = \left.N_{\text{CE}\nu\text{NS}}\right|_{\rm SM} = 44.9$ and $\mathcal{E} = 6~ {\rm tonne} \cdot {\rm year} $. 
We consider two cases with the total systematic uncertainty $\sigma_\alpha =0.245$ being included or not. The resulting constraints are represented in lighter red and darker red bands with solid and dashed boundaries, respectively. One can observe that the anticipated constraint from PandaX-4T with the systematic uncertainties is weaker than that from COHERENT.
If in the future the systematic uncertainties are well controlled, with the assumption that they are negligible, we could achieve a sensitivity comparable to that of the COHERENT CsI measurement.

\section{Conclusion}
\label{sec:conclusion}

In this work, we have studied the constraints on the neutrino non-standard interactions (NSIs) using the latest \cevns\ measurements in the COHERENT and PandaX-4T and XENONnT experiments.
The cross section of \cevns\ is calculated within an end-to-end effective field theory framework. In this approach, the dimension-6 operators in the low-energy EFT are matched to hadronic operators in the heavy baryon chiral perturbation theory, while the nuclear response of target nuclei is described using multipole expansions.

We have performed the $\chi^2$ analyses of the \cevns\ events observed in the CsI[Na] detector of COHERENT and PandaX-4T and XENONnT, where both the impacts of neutrino NSIs on the solar matter effect in the propagation of solar neutrinos and the \cevns\ cross section in the detection are considered. We have derived the one-parameter-a-time lower bounds on the NSI scale for specific neutrino flavors, and we have obtained that the constraints for the $e$ and $\mu$ flavors from COHERENT are more stringent than those from PandaX-4T and XENONnT. Nevertheless, the latter can provide unique sensitivities to the neutrino NSIs for the $\tau$ flavor due to the oscillation of solar $^8$B neutrinos propagating from the Sun to the Earth.

Besides, we have compared the two-dimensional constraints on the NSI parameters from COHERENT, PandaX-4T and XENONnT, and have identified that the expected number of \cevns\ is more sensitive to neutrino NSIs appearing in the cross section compared to averaged probability of solar neutrinos.

Moreover, we have found that the sensitivity of PandaX-4T is limited by the central value of measured \cevns\ counts.  To make projections of the future prospects, we have assessed the measurements of the solar $^8$B neutrinos via \cevns\ in dark matter detectors, leveraging the current results from PandaX-4T as a reference.
Assuming that the central value of measured \cevns\ counts aligns with the SM prediction, the sensitivity is significantly improved for the exposure of  $ 6~ {\rm tonne} \cdot {\rm year} $, and is comparable to that imposed by the COHERENT CsI measurement if the systematic uncertainties are further disregarded.

{\bf Note added:} After this paper was finished, another paper~\cite{AristizabalSierra:2024nwf} appeared, which has some overlap of our work. However, we use an end-to-end EFT framework to investigate the neutrino NSIs sensitivities, and we emphasize that it is important to include systematic uncertainties in the analysis.


\acknowledgments
We would like to express our gratitude to Yu-Feng Li for the valuable help with the $\chi^2$ fitting of the COHERENT data. GL also thanks Xun-Jie Xu for useful correspondence regarding Ref.~\cite{Lindner:2016wff}. FT thanks Ningqiang Song for the discussion on the solar neutrino and Bing-Long Zhang for the discussion on numerical calculations. This work is supported by the National Science Foundation of China under Grants No. 12347105,
No. 12375099 and No. 12047503, and the National Key Research and Development Program of China
Grant No. 2020YFC2201501, No. 2021YFA0718304. GL is also supported by the Guangdong Basic and Applied Basic Research Foundation (2024A1515012668), and SYSU startup funding.

\appendix

\section{Detector resolution and efficiency in COHERENT CsI measurement}
\label{app:coherent}

In Eq.~\eqref{eq:event_coherent}, the number of PEs is~\cite{Papoulias:2019txv,COHERENT:2021xmm}
\begin{align}
    n_{\mathrm{PE}} = 13.35~ \dfrac{T_{ee}}{\rm keV} \;,
\end{align}
where the true electron-equivalent recoil energy $T_{ee}$ is related to the true nuclear recoil energy as
\begin{align}
    T_{ee} = f_Q(T_{\rm nr}) T_{\rm nr}\;.
\end{align}
The quenching factor $f_Q(T_{\rm nr})$ can be parameterized as~\cite{COHERENT:2021pcd} 
\begin{align}
    f_Q(T_{\rm nr}) = k_0 + k_1 T_{\rm nr} + k_2 T_{\rm nr}^{2} + k_3 T_{\rm nr}^{3}\;,
\end{align}
where the parameters $k_0=0.05546$, $k_1=4.307$, $k_2=-111.7$ and $k_3=840.4$. 

The detector energy resolution $P(n_{\rm PE})$ is modeled with the gamma function,
\begin{align}
   P(n_{\rm PE})=\frac{(a(1+b))^{1+b}}{\Gamma(1+b)} n_{\rm PE}^b e^{-a(1+b) n_{\rm PE}}\;,
\end{align}
where $a=0.0749{~\rm keV}/E_{ee}$ and $b=9.56E_{ee}/{\rm keV}$. 

The reconstructed energy and time are uncorrelated, thus allowing us to deal with the energy and time efficiency independently~\cite{COHERENT:2021xmm},
\begin{align}
    \varepsilon_E(n_{\rm PE})&=\frac{a_1}{1+e^{-b_1(n_{\rm PE}-c_1)}}+d_1\;,\\
    \varepsilon_T\left(t_{\rm r e c}\right)&= \begin{cases}1\;, & t_{\rm r e c}<a_2\;, \\ e^{-b_2\left(t_{\rm r e c}-a_2\right)}\;, & t_{\rm r e c} \geq a_2\;,\end{cases}    
\end{align}
where the parameters are $a_1=1.32$,  $b_1=0.285$, $c_1=10.9$, $d_1=-0.333$, and $a_2=0.52~\mu$s, $b_2= 0.0494/\mu$s~\cite{COHERENT:2021xmm}.

The efficiency $\varepsilon(n_{\rm PE})$ has been implemented in the integration over $n_{\rm PE}$ in Eq.~\eqref{eq:event_coherent}. On the other hand, we consider the average time efficiency
\begin{align}
    \langle \varepsilon_{T} \rangle_{\nu_\alpha} \equiv \dfrac{\sum_{j} \int_{t_{\rm rec}^i}^{t_{\rm rec}^{j+1}} {\rm d} t_{\rm rec} ~\varepsilon_{T}(t_{\rm rec}) N^{j}_{\nu_\alpha}(t_{\rm rec})}{\sum_{j} \int_{t_{\rm rec}^i}^{t_{\rm rec}^{j+1}} {\rm d} t_{\rm rec} N^{j}_{\nu_\alpha}(t_{\rm rec})}\;,
\end{align}
where $N^{j}_{\nu_\alpha}(t_{\rm rec})$ represents the $n_{\rm PE}$-integrated expected number of the SM \cevns\ events in the reconstructed time $t_{\rm rec}$ for each flavor of neutrino flux depicted in the right panel of Figure 1 in Ref.~\cite{COHERENT:2021xmm}. Due to variations in neutrino arrival times, 
the time efficiencies for different flavors are distinct, $\langle \epsilon_{T} \rangle_{\nu_\mu}=0.994$, $\langle \epsilon_{T} \rangle_{\bar{\nu}_\mu}=0.918$ and $\langle \epsilon_{T} \rangle_{\nu_e}=0.92$.

\section{The impact of neutrino NSIs on the propagation}
\label{app:propagation}

During the propagation of solar neutrinos to the Earth, the matter effect is significantly affected by matter effects on the averaged probabilities $\langle P_{\nu_\alpha}\rangle$. In the presence of neutrino NSIs, the solar matter effect would be modified. In this appendix, we will provide the details of the averaged probability with the impact of neutrino NSIs being included. 

There are two stages for the solar neutrino state propagating from the Sun to the detector:
propagation in the Sun and in the vacuum. In the Sun, we need to consider the matter effect  \cite{deHolanda:2004fd, Maltoni:2015kca}, 
\begin{equation}
    \mid \nu_{e} \rangle = \sum_i U_{ei}^m \mid \nu_{im} \rangle \;,
\end{equation}
where $U^m$ and $\mid \nu_{m} \rangle$ are the mixing matrix and eigenstate of the Hamiltonian in matter. Because of the small density gradient, we can use the adiabatic approximation  \cite{Mikheev:1986if, Bethe:1986ej}, which means the different eigenstates evolve independently in the Sun \cite{Maltoni:2015kca}.

By considering the neutrino NSIs, the Hamiltonian in the matter can be expressed as
\begin{eqnarray}
&  & H= 
\left(\begin{array}{ccc}
\Delta H_1  \\
& \Delta H_2\\
&  & \Delta H_3
\end{array}\right)  \nonumber
+
\sqrt{2} G_F n_e \begin{pmatrix}
1 &  & \\
 & 0 & \\
 & & 0 \\
\end{pmatrix}
\\
&  &\quad \quad  + 
\sqrt{2} G_F n_{qV} \begin{pmatrix}
\epsilon_{ee}^{qV} & \epsilon_{e\mu}^{qV} & \epsilon_{e\tau}^{qV}\\
\epsilon_{e\mu}^{qV*} & \epsilon_{\mu\mu}^{qV} & \epsilon_{\mu\tau}\\
\epsilon_{e\tau}^{qV*} & \epsilon_{\mu\tau}^{qV*} & \epsilon_{\tau\tau}^{qV}
\end{pmatrix}
\,,
\end{eqnarray}
where $ \Delta H_i \equiv U (m_{i}^{2}/{2E_{\nu}}) U^{\dagger}$, $ i = 1,2,3$, $U$ denotes the Pontecorvo-Maki-Nakagawa-Sakata (PMNS) matrix, and $n_q (q=u,d)$ is the number density of the quark inside the Sun. The NSI parameters, $\epsilon_{\alpha \beta}^{qV} = (\epsilon_{\alpha \beta}^{L} +\epsilon_{\alpha \beta}^{R} )$, which are related to the Wilson coefficients via Eq.~\eqref{eq:epsLR}.
In the vacuum, there is no matter effect, thus the Hamiltonian is in the special case of $n_e=0$ and $n_q =0$. 
As to the matter effect of propagating in the Earth, it can be neglected compared to the propagation in the Sun, the probability from the Sun to the Earth is then
\begin{equation} 
    P_{\nu_{\alpha}} = |\langle \nu_e(t_0) | \nu_{\alpha}(t_E) \rangle|^2
    = \sum_i |U^m_{ei}|^2 |U_{\alpha i}|^2 \,,
\end{equation}
where $i=1,2,3$ depicts the propagation state, while the flavor index $\alpha = e, \mu, \tau$. The wave packets of different neutrino eigenstates with different group velocities are separated in the propagation \cite{Maltoni:2015kca}, so that different neutrino states are incoherent.

For simplicity, we make the two-flavor approximation, and the Hamiltonian can be written as~\cite{Friedland:2004pp,Gonzalez-Garcia:2013usa,Esteban:2018ppq,Coloma:2022umy,Amaral:2023tbs}
\begin{align}
  \label{eq:2times2-ham}
  H=\frac{1}{4E_\nu}
  \begin{pmatrix}
    -\Delta m_{21}^2 \cos2\theta_{12} + A 
    & \Delta m_{21}^2 \sin2\theta_{12} + B\\
    \Delta m_{21}^2 \sin2\theta_{12} + B 
    & \Delta m_{21}^2 \cos2\theta_{12} - A
  \end{pmatrix} \,
\end{align}
with
\begin{align}
    A(r, E_{\nu}, \varepsilon_D^q) 
    &= 4\sqrt{2}E_\nu G_F n_e(r) \left[\frac{\cos^2\theta_{13}}{2} - \frac{n_q(r)}{n_e(r)} \varepsilon_D^q\right]
    \ , \nonumber\\
    B(r, E_{\nu}, \varepsilon_N^q) &=4\sqrt{2}E_\nu G_F n_q(r) \varepsilon_N^q \ ,
\end{align}
where quark number densities $ n_u = 2 n_e + n_n $, $ n_d = n_e + 2n_n $, and $r$ is defined as the ratio of the distance to the center of the Sun to the Sun's radius.

For the CE$\nu$NS experiments, we concentrate on the diagonal terms $\epsilon_{ee}^{qV}$, $\epsilon_{\mu \mu}^{qV}$, $\epsilon_{\tau \tau}^{qV}$, 
\begin{align}
  \varepsilon_D^q&=
  -\frac{c_{13}^2}{2}\epsilon_{ee}^{qV}
  +\frac{\left[c_{13}^2 - \left(s_{23}^2 - s_{13}^2 c_{23}^2\right)\right]}{2}
  \epsilon_{\mu\mu}^{qV} \nonumber \\
  & \qquad +\frac{\left(s_{23}^2 - c_{23}^2s_{13}^2\right)}{2}
  \epsilon_{\tau\tau}^{qV} \,, \nonumber \\
  \varepsilon_N^q&= -s_{13}c_{23}s_{23}\epsilon_{\mu\mu}^{qV}
  + s_{13}c_{23}s_{23}\epsilon_{
\tau\tau}^{qV} \,,
\end{align}
where $c_{ij}\equiv \cos\theta_{ij}$, and $s_{ij}\equiv \sin\theta_{ij}$.

The survival probability $P_{\nu_e}$ in the two-flavor approximation is expressed as \cite{Friedland:2004pp,Gonzalez-Garcia:2013usa, Xu:2022wcq}
\begin{align}
  P_{\nu_e}(E_\nu,r)=c_{13}^4
  \, 
  ({1+\cos2\theta^m_{12}(r)\cos2\theta_{12}})/2  + s_{13}^4 \,,
\end{align}
where 
\begin{equation} \label{eq.theta12m}
  \cos2\theta^m_{12}(r) = \frac{\Delta m_{12}^2\cos2\theta_{12}-A}
  {\sqrt{\left(\Delta m_{12}^2\cos2\theta_{12}-A\right)^2
      +
  \left(\Delta m_{12}^2\sin2\theta_{12} +B \right)^2}}\;.
\end{equation}

The survival probability averaged over the $^8$B neutrino flux is
\begin{equation}
  \langle P_{\nu_e}(E_\nu) \rangle = \frac{\int_{0}^{1} dr \rho(r) P_{\nu_e}(E_\nu,r)}{\int_{0}^{1} dr \rho(r)} \,,
\end{equation}
where the $\rho(r)$ is the number density distribution of $^8$B neutrino production. 

For the $P_{\nu_{\mu}}$ and $P_{\nu_{\tau}}$, we need to calculate the three-flavor mixing matrix $U_{ei}^{m}$, which relates to $\theta_{12}^{m}$ and $\theta_{13}^{m}$. We make the assumption that the $\theta_{12}^{m}$ is similar to its value in the two-flavor approximation in Eq.~\eqref{eq.theta12m}, and $\theta_{13}^{m}$ is sufficiently small and approximately equal to the $\theta_{13}$ in the vacuum.


\bibliographystyle{apsrev4-1}
\bibliography{reference.bib}

\end{document}